\documentclass[12pt]{iopart}

\newcommand{\dd}{\,\mathrm{d}}

\usepackage{bm}
\usepackage{iopams}
\usepackage{braket}
\usepackage{setstack}
\usepackage{citesort}
\usepackage{cite}
\usepackage{longtable}
\usepackage{caption}
\usepackage{multirow}
\usepackage{graphicx}
\usepackage{ntheorem}
\usepackage{stackrel}
\usepackage{subfigure}
\usepackage{hhline}
\usepackage{epstopdf}
\usepackage[fleqn]{cases}

\newtheorem{proposition}{Proposition}
\newtheorem*{conjecture}{Conjecture}

\begin{document}
\title{Second-order statistics of fermionic Gaussian states}
\author{Youyi Huang and Lu Wei}
\address{Department of Computer Science, Texas Tech University, Texas 79409, USA}
\ead{\{youhuang,luwei\}@ttu.edu}
\vspace{10pt}
\begin{indented}
\item[]November 2021
\end{indented}

\begin{abstract}
We study the statistical behavior of entanglement in quantum bipartite systems over fermionic Gaussian states as measured by von Neumann entropy and entanglement capacity. The focus is on the variance of von Neumann entropy and the mean entanglement capacity that belong to the so-defined second-order statistics. The main results are the exact yet explicit formulas of the two considered second-order statistics for fixed subsystem dimension differences. We also conjecture the exact variance of von Neumann entropy valid for arbitrary subsystem dimensions. Based on the obtained results, we analytically study the numerically observed phenomena of Gaussianity of von Neumann entropy and linear growth of average capacity.
\end{abstract}

\vspace{2pc}

\maketitle
\section{Introduction}\label{sec:intr}
Quantum information theory aims to construct the theoretical groundwork of quantum technologies such as quantum computing and quantum communications. In the exploitation of the revolutionary advances of quantum mechanical systems, the understanding of the non-classical phenomenon of quantum entanglement is quite crucial. Quantum entanglement is also the resource and medium that enable quantum technologies.

In this work, we study the statistical behavior of entanglement of quantum bipartite systems of fermionic Gaussian states. In the literature, the degree of entanglement as measured by entanglement entropies such as von Neumann entropy, quantum purity, and Tsallis entropy has been studied over different ensembles, e.g. the Hilbert-Schmidt ensemble~\cite{Lubkin78, Page93,MML02,Foong94,Ruiz95,Giraud07,VPO16,Wei17,Wei19T, Wei20, HWC21} and the Bures-Hall ensemble~\cite{Sommers04,Osipov10,Borot12,SK2019, Wei20BHA, Wei20BH, LW21}. Similar investigations are now being carried out over the fermionic Gaussian ensemble, where the average von Neumann entropy is obtained in~\cite{BHK21}. As an important step towards further understanding its statistical distribution, we aim to derive the variance. The variance describes the fluctuation of the entropy around its mean value. It also provides crucial information such as if the average entropy is typical. In addition, we study the capacity of entanglement over the fermionic Gaussian ensemble. Entanglement capacity is another entanglement measure that possesses other distinctive properties as compared to entanglement entropies~\cite{BNP21}.

We now introduce the formulation that leads to the fermionic Gaussian states as follows. Consider a phase space $V$ in $\mathbb{R}^{2N}$ given by a $2N$-dimensional real vector space that corresponds to a system of $N$ fermionic degrees of freedom. Its dual space $V^{*}$ in $\mathbb{R}^{2N}$ and $V$ are equipped with a positive-definite bilinear form $G^{ab}:V^{*}\times V^{*}\to\mathbb{R}$ and its inverse $G_{ab}^{-1}: V\times V\to\mathbb{R}$. Observables in quantum theory are linear operators on a Hilbert space. Here, it is convenient to introduce an operator-valued vector $\hat{\xi}^{a}$, i.e., a quantization map~\cite{HJ19,HB21}
\begin{equation}\label{eq}
\{\hat{\xi}^{a},\hat{\xi}^{b}\}=G^{ab}\mathbb{I}
\end{equation}
with $\{\hat{\xi}^{a},\hat{\xi}^{b}\}=\hat{\xi}^{a}\hat{\xi}^{b}+\hat{\xi}^{b}\hat{\xi}^{a}$ denoting the anti-commutation relation and $\mathbb{I}$ being an identity operator. The operators $\hat{\xi}^{a}$ can be represented via the bases of the Majorana operators or fermionic creation and annihilation operators as discussed in details in~\cite{HB21}. A fermionic Gaussian state is labeled by its covariance matrix
\begin{equation}\label{eq}
\Omega^{ab}=\Bra{\Omega}\hat{\xi}^{a}\hat{\xi}^{b}-\hat{\xi}^{b}\hat{\xi}^{a}\Ket{\Omega},
\end{equation}
which is clearly anti-symmetric.

Given a fermionic Gaussian state, the phase space $V$ of a system of $N$ fermionic degrees of freedom can be decomposed into two orthogonal complementary subsystems $A$ and $B$ of the dimensions $m$ and $n$, respectively,
\begin{equation}\label{eq}
V=A\oplus B.
\end{equation}
Here, $m+n=N$ and we assume $m\leq n$ without loss of generality. One can always choose an orthonormal basis $\left(q_1^{A},p_1^{A},\dots,q_m^{A},p_m^{A},q_1^{B},p_1^{B},\dots,q_n^{B},p_n^{B}\right)$ such that the covariance matrix is transformed into the form
\begin{equation}\label{eq}
\Omega=\left(\begin{array}{cc}
\Omega_{A} & \Omega_{AB} \\
\Omega_{BA} & \Omega_{B} \\
\end{array}\right).
\end{equation}
In the above decomposition, the mode $(q_i^{A},p_i^{A})$ is entangled with the mode $(q_i^{B},p_i^{B})$ for each $i=1,\dots,m$ and the $2m\times 2m$ subsystem $\Omega_{A}$ and the $2n\times 2n$ subsystem $\Omega_{B}$ have the same spectrum except for the $2n-2m$ eigenvalues of the values $\pm \imath$, $\imath=\sqrt{-1}$. Namely, $\Omega_{A}$ and $\Omega_{B}$ contain the same information regarding the entanglement between the two subsystems. The matrix $\Omega_{A}$ takes the block diagonal form~\cite{HJ19,HB21}
\begin{equation}\label{eq}
\Omega_{A}=\left(\begin{array}{ccc}
\cos(2r_1)\mathbb{A} & \dots & 0 \\
\vdots & \ddots & \vdots \\
0 & \dots & \cos(2r_m)\mathbb{A} \\
\end{array}\right),
\end{equation}
where
\begin{equation}\label{eq}
\mathbb{A}=\left(\begin{array}{cc}
0 & 1 \\
-1 & 0 \\
\end{array}\right)
\end{equation}
and the amount of entanglement is now encoded in the parameters $r_{i}\in[0,\pi/4]$. Therefore, by defining the new variables $x_i=\cos(2r_i)$, $i=1,\dots,m$, the von Neumann entropy of the considered fermionic system can be represented as~\cite{BHK21}
\begin{eqnarray}\label{eq:vN}
S&=&-\tr\left(\frac{\mathrm{I}_m+\imath\Omega_{A}}{2}\ln\frac{\mathrm{I}_m+\imath\Omega_{A}}{2}\right) \\
&=&-\sum_{i=1}^{m}v(x_{i}), \qquad x_{i}\in[0,1], \label{eq:vN2}
\end{eqnarray}
where
\begin{equation}\label{eq:vx}
v(x)=\frac{1-x}{2}\ln\frac{1-x}{2}+\frac{1+x}{2}\ln\frac{1+x}{2},
\end{equation}
and $\mathrm{I}_m$ is an identity matrix of dimension $m$.
The density of the eigenvalues $x_i, i=1,\dots,m$, follows a non-standard Jacobi unitary ensemble, which is proportional to~\cite{BHK21}
\begin{equation}\label{eq:nJ}
\prod_{1\leq i<j\leq m}\left(x_{i}^2-x_{j}^2\right)^{2}\prod_{i=1}^{m}\left(1-x_{i}^2\right)^{n-m}.
\end{equation}
The above random matrix ensemble is referred to as the fermionic Gaussian ensemble.  Its eigenvalues are supported in
\begin{equation}
x_{i}\in[0,1]
\end{equation}
instead of the natural support $x_{i}\in[-1,1]$ of a standard Jacobi ensemble. This corresponds to a bipartite quantum system in the fermionic Gaussian states of subsystem dimensions $m$ and $n$ with the assumption $m\leq n$.
 For convenience, we denote
\begin{equation}\label{eq:a}
a=n-m
\end{equation}
as the difference of subsystem dimensions.

We define the $k$-th order statistic over the fermionic Gaussian ensemble as the average of random variables consisting of sum of all eigenvalues $x_i, i=1,\dots,m$ of the form
\begin{equation}\label{eq:kths}
\ln^{k}(1\pm x_i).
\end{equation}
In this work, we focus on the study of two major second-order statistics of fermionic Gaussian states. The first one is the variance of the von Neumann entanglement entropy~\eref{eq:vN2}.
The von Neumann entropy is the most fundamental entanglement measure that satisfies various interesting properties. It is an important theoretic quantity in quantum information theory and has applications in different fields, including the many-body systems and black hole radiation~\cite{BZ06}. Recently, the average von Neumann entropy over the fermionic Gaussian ensemble~\eref{eq:nJ} has been obtained as~\cite{BHK21}
\begin{eqnarray}
\mathbb{E}\!\left[S\right]&=&\left(m+n-\frac{1}{2}\right)\psi_{0}(2m+2n)+\left(\frac{1}{4}-m\right)\psi_{0}(m+n)+\nonumber\\
&&\left(\frac{1}{2}-n\right)\psi_{0}(2n)-\frac{1}{4}\psi_{0}(n)-m\label{eq:m1},
\end{eqnarray}
where
\begin{equation}\label{eq:pl}
\psi_{0}(x)=\frac{\dd\ln\Gamma(x)}{\dd x}
\end{equation}
is the digamma function and for a positive integer~\cite{Brychkov08}
 \begin{equation}\label{eq:pl0}
\psi_{0}(l)=-\gamma+\sum_{k=1}^{l-1}\frac{1}{k}
\end{equation}
with $\gamma\approx0.5772$ being the Euler's constant. As one step further towards understanding the statistical behaviour of von Neumann entropy, we study the exact variance $\mathbb{V}\!\left[S\right]$ over the fermionic Gaussian states. Specifically, we are able to derive the exact variance for equal subsystem dimensions $n=m$. We also conjecture the exact variance for arbitrary subsystem dimensions $n\geq m$.

Another second-order statistic we are considering is the mean value of the entanglement capacity. The capacity of entanglement over the fermionic Gaussian ensemble~\eref{eq:nJ} is defined as~\cite{Nandy,BNP21}
\begin{equation}\label{eq:Cac}
C=\sum_{i=1}^{m}\frac{1-x_{i}^{2}}{4}\ln^2\frac{1+x_i}{1-x_i}.
\end{equation}
The mean capacity is a second-order statistics because the above definition can be rewritten as
\begin{eqnarray}\label{eq:capacity}
C&=&\sum_{i=1}^{m}\Bigg(\frac{1-x_i}{2}\ln^2\frac{1-x_i}{2}+\frac{1+x_i}{2}\ln^2\frac{1+x_i}{2}-\nonumber\\
&&\left(\frac{1-x_i}{2}\ln\frac{1-x_i}{2}+\frac{1+x_i}{2}\ln\frac{1+x_i}{2}\right)^2\Bigg),
\end{eqnarray}
which corresponds to our definition of $k$-th order statistic~\eref{eq:kths} when $k=2$.
As the quantum information theoretic counterpart of heat
capacity, the capacity of entanglement is known to capture the deviation from the maximal entanglement~\cite{Nandy, Boer19}. The mean of entanglement capacity under the Hilbert-Schmidt ensemble has been recently computed in~\cite{OKUYAMA21}. For the fermionic Gaussian ensemble, we compute the exact average capacity when the subsystem dimension differences are $a=0,1,2,3$.


In proposition \ref{prop1} below, we summarize the exact variance of von Neumann entanglement entropy of equal subsystem dimensions $n=m$.
\begin{proposition}
For equal subsystem dimensions $n=m$, i.e., $a=0$, the exact variance of the  von Neumann entropy~\eref{eq:vN2} of fermionic Gaussian states~\eref{eq:nJ} is given by
\begin{eqnarray}\label{prop1}
 \fl \mathbb{V}\!\left[S\right]&=&\!\left(\frac{1}{2}-2 n\right) \psi _1(4 n)+\frac{56 n^2-36 n+5}{8(4 n-1)}\psi _1(2 n)+\frac{\psi _1(n)}{8}-\frac{ \psi_0(4 n)}{2}+\frac{ \psi_0(2 n)}{2},\label{eq:prop1}
\end{eqnarray}
where $\psi_{0}(x)$ is the digamma function~\eref{eq:pl} and $\psi_{1}(x)=\dd^{2}\ln\Gamma(x)/\dd x^{2}$ is the trigamma function.
\end{proposition}\label{prop1}
The proof of proposition~\ref{prop1} can be found in~\sref{sec:2.2}. Here, the trigamma function also admits the finite sum form as~\cite{Brychkov08}
\begin{equation}\label{eq:pl1}
\psi_{1}(l)=\frac{\pi^{2}}{6}-\sum_{k=1}^{l-1}\frac{1}{k^{2}}.
\end{equation}
Our approach in deriving the result of proposition~\ref{prop1} in principle works for any fixed dimension difference (i.e., $a=n-m=1,2,\dots$). Based on the further calculations of the variance for the cases of different values of $a$, we come up with the following conjecture on the exact variance of arbitrary subsystem dimensions.
\begin{conjecture}
For any subsystem dimensions $m\leq n$, the exact variance of the von Neumann entanglement entropy~\eref{eq:vN2} of fermionic Gaussian states~\eref{eq:nJ} is given by
 \begin{eqnarray}\label{eq:conj}
 \fl \mathbb{V}\!\left[S\right]&=&\!\left(\frac{1}{2}-m-n\right) \psi _1(2 m+2 n)+\left(n-\frac{1}{2}\right) \psi _1(2 n)+\left(\frac{m (2 m+n-1)}{2 m+2 n-1}-\frac{1}{8}\right)\times\nonumber\\
\fl && \psi _1(m+n)+\frac{\psi _1(n)}{8}-\frac{1}{2} (\psi_0(2 m+2 n)-\psi_0(2 n)).
\end{eqnarray}
\end{conjecture}
Note that in the case of equal subsystem dimensions $n=m$, the above conjectured result reduces to~\eref{eq:prop1} in proposition~\ref{prop1}, as expected.  It is also worth mentioning that the conjectured formula~\eref{eq:conj} recovers the obtained asymptotic variance in~\cite{BHK21}, i.e.,
\begin{equation}\label{eq:valim}
\lim_{m,n \rightarrow \infty}\mathbb{V}\!\left[S\right]=\frac{1}{2}\left(f+f^2+\ln (1-f)\right)
\end{equation}
for a fixed $f=m/(n+m)$. This asymptotic result~\eref{eq:valim} can be obtained by using the asymptotic behavior of polygamma functions~\cite{Brychkov08}
\begin{eqnarray}
\psi_0(x)&=&\Theta\left(\ln (x)-\frac{1}{2 x}\right),\qquad x\to\infty \label{eq:limpl0}\\
\psi_{j}(x)&=&\Theta\left(\frac{1}{x^{j}}\right),\qquad x\to\infty,~~~~~~j\geq1,\label{eq:limpl1}
\end{eqnarray}
where $\Theta(\cdot)$ is the Big-Theta notation of the Bachmann-Landau symbols.

We now move on to the results of the entanglement capacity. Here, we also focus on cases when the differences of the subsystem dimensions are fixed.
The exact formulas of mean capacity of subsystem dimension differences $a=n-m=0,1,2,3$ are summarized in proposition~\ref{prop2} below, where the corresponding calculation is provided in~\sref{sec:2.3}.
\begin{proposition}\label{prop2}
For subsystem dimension differences $a=n-m=0,1,2,3$, the respective averages of entanglement capacity~\eref{eq:Cac} of fermionic Gaussian states~\eref{eq:nJ} are
\begin{eqnarray}
 \mathbb{E}\!\left[C\right]&=&a_0 \psi _1(2 n)+a_1 \psi _1(n)+a_2 \label{eq:prop20}\\
 \mathbb{E}\!\left[C\right]&=&b_0 \psi _1(2 n)+b_1 \psi _1(n)+b_2\label{eq:prop21}\\
 \mathbb{E}\!\left[C\right]&=&c_0 \psi _1(2 n)+c_1 \psi _1(n)+c_2 (\psi_0 (2 n)-\psi_0 (1))+c_3\label{eq:prop22}\\
 \mathbb{E}\!\left[C\right]&=&d_0 \psi _1(2 n)+\!d_1 \psi _1(n)+\!d_2 (\psi_0 (2 n)-\psi_0 (1))+d_3\label{eq:prop23},
\end{eqnarray}
where the coefficients $a$, $b$, $c$, and $d$ are summarized in~\tref{table1}.
\end{proposition}

\begin{table}[htbp]
\centering
\normalsize
\caption{Coefficients of $\mathbb{E}\!\left[C\right]$ of subsystem dimension differences $a=0,1,2,3$} \label{table1}
\renewcommand\thesubtable{}
\begin{subtable}[\label{table:1a} $a=0$]
{
\begin{tabular}{l  p{13.5cm}}
\hhline{==}
\bs\multirow{1}{*}{$a_0=$} & $\displaystyle -\frac{(2 n-1)^2}{2 (4 n-1)}$  \\
\bs\multirow{1}{*}{$a_1=$} & $\displaystyle -\frac{1}{8}$  \\
\bs\multirow{1}{*}{$a_2=$} & $\displaystyle \frac{\pi ^2 \left(8 n^2-4 n+1\right)}{16 (4 n-1)}+\frac{1}{2} (1-2 n)$  \\
\bs
\end{tabular}}
\end{subtable}
\begin{subtable}[\label{table:1b} $a=1$]
{
\begin{tabular}{l  p{13.5cm}}
\hhline{==}
\bs
\multirow{1}{*}{$b_0=$} & $\displaystyle -\frac{4 n^2-8 n+3}{2 (4 n-3)}$  \\
\bs\multirow{1}{*}{$b_1=$} & $\displaystyle -\frac{1}{8}$  \\
\bs\multirow{1}{*}{$b_2=$} & $\displaystyle \frac{\pi ^2 \left(8 n^2-12 n+3\right)}{16 (4 n-3)}-\frac{16 n^3-36 n^2+28 n-9}{2 (2 n-1) (4 n-3)}$  \\
\bs
\end{tabular}}
\end{subtable}
\end{table}
\pagebreak
\normalsize
\begin{table}[htbp]
\renewcommand\thesubtable{}
\ContinuedFloat
\begin{subtable}[\label{table:1c} $a=2$]
{
\begin{tabular}{l  p{13.5cm}}
\hhline{==}
\bs\multirow{1}{*}{$c_0=$} & $\displaystyle \frac{-4 n^2+12 n-5}{8 n-10}$  \\
\bs\multirow{1}{*}{$c_1=$} & $\displaystyle -\frac{1}{8} $  \\
\bs\multirow{1}{*}{$c_2=$} & $\displaystyle \frac{1}{2 n^2-5 n+3}$  \\
\bs\multirow{1}{*}{$c_3=$} & $\displaystyle \frac{\pi ^2 \left(8 n^2-20 n+5\right)}{64 n-80}-\frac{32 n^4-152 n^3+268 n^2-210 n+75}{2 (2 n-3) (2 n-1) (4 n-5)}$  \\
\end{tabular}}
\end{subtable}

\begin{subtable}[\label{table:1d} $a=3$]
{
\begin{tabular}{l  p{13.5cm}}
\hhline{==}
\bs\multirow{1}{*}{$d_0=$} & $\displaystyle -\frac{(2 n-7) (2 n-1)}{2 (4 n-7)}$  \\
\bs\multirow{1}{*}{$d_1=$} & $\displaystyle -\frac{1}{8}$  \\
\bs\multirow{1}{*}{$d_2=$} & $\displaystyle \frac{2 \left(4 n^2-14 n+11\right)}{(n-2) (n-1) (2 n-5) (2 n-3)}$  \\
\bs\multirow{1}{*}{$d_3=$} & $\displaystyle \frac{\pi ^2 \left(8 n^2-28 n+7\right)}{16 (4 n-7)}-\frac{1}{2 (n-2) (n-1) (2 n-5) (2 n-3) (2 n-1) (4 n-7)}\times$  \\
\bs\multirow{1}{*}&$\displaystyle(64 n^7-720 n^6+3408 n^5-8736 n^4+13176 n^3-11967 n^2+6258 n-1470)$\\
\bs
\hhline{==}

\end{tabular}}

\end{subtable}
\end{table}
From the results of proposition \ref{prop2}, we observe that the coefficients of $\psi_1(n)$ remain the same for $a=0,1,2,3$. We also observe that the term involving digamma functions, i.e., $(\psi_0 (2 n)-\psi_0 (1))$, starts to appear when $a\geqslant2$. Our approach in deriving the above formulas in principle works for any given subsystem dimension difference $a=n-m$. Note that in a recent work~\cite{BNP21}, an exact representation of mean capacity for arbitrary $a$ has been obtained in terms of a rather complicated triple summations.

As will be seen in~\sref{sec:2}, the  computation of the exact second-order statistics of the  fermionic Gaussian states requires one to calculate the corresponding integrals involving Jacobi polynomials. In principle, one could obtain different representations of the same integral due to the availability of different summation representations such as the ones of Jacobi polynomial~\eref{eq:J1}--\eref{eq:J3}. These different choices lead to different summation forms, where most of the combinations may not permit further simplifications. In practice, the integral identities utilized in establishing the claimed results have underwent a sensible choice so as to facilitate the cancellation among the summations. The cancellation eventually leads to the desired closed-form results as shown in proposition \ref{prop1} and proposition \ref{prop2}.

Finally, we point out the fact that our obtained summation representations~\eref{eq:A1S}--\eref{eq:ICS} may not seem to permit further simplification that would lead to the conjectured variance~\eref{eq:conj} for an arbitrary $a$. This is because the gamma functions involved in these summation representations, e.g. in~\eref{eq:A2S} and~\eref{eq:B2S}, will not cancel in pairs. In contrast, for a given $a$, these gamma functions are reduced to summable rational terms leading to the claimed results.

\begin{figure}[!h]
\centering
\includegraphics[width=0.85\linewidth]{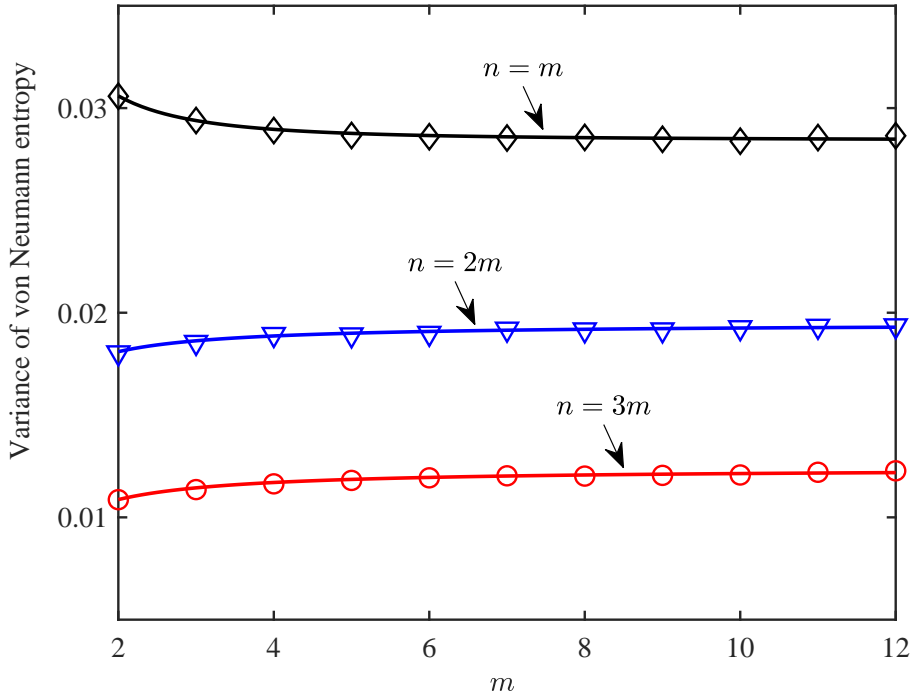}
\caption{Variance of von Neumann entropy: analytical results versus simulations. The solid black line represents the obtained analytical result~\eref{eq:prop1} of equal subsystem dimensions $n=m$. The solid blue line ($n=2m$) and the solid red line ($n=3m$) are drawn by the conjectured result~\eref{eq:conj}. The scatters (diamond, triangular, and square shapes) represent numerical simulations. }
\label{fig:p1}
\end{figure}

To illustrate the derived result~\eref{eq:prop1} and provide numerical evidence of the conjecture~\eref{eq:conj}, we plot in~\fref{fig:p1}\footnote{The simulations performed in~\fref{fig:p1}--3 make use of the Mathematica codes provided by Santosh Kumar based on the log-gas approach as discussed in\cite[appendix B]{SK2019}.} the exact variance of von Neumann entropy as compared with the simulations. We consider the scenarios of equal and unequal subsystem dimensions, where the claimed results in solid lines match well with the corresponding simulations as represented by the scatters. We also observe that the slope of the curves approaches to zero when the dimension of the system becomes larger. This phenomenon can be explained by the asymptotic formula~\eref{eq:valim} of the variance, i.e., the variance is dominated by the ratios of subsystem dimensions for large-dimensional systems.

We now consider the standardized von Neumann entropy
\begin{equation}\label{eq:X}
X=\frac{S-\mathbb{E}\!\left[S\right]}{\sqrt{\mathbb{V}\!\left[S\right]}},
\end{equation}
where its average $\mathbb{E}\!\left[S\right]$ and variance $\mathbb{V}\!\left[S\right]$ are given by~\eref{eq:m1} and~\eref{eq:conj}, respectively.
\pagebreak
\begin{figure}[!h]
\centering
\includegraphics[width=0.85\linewidth]{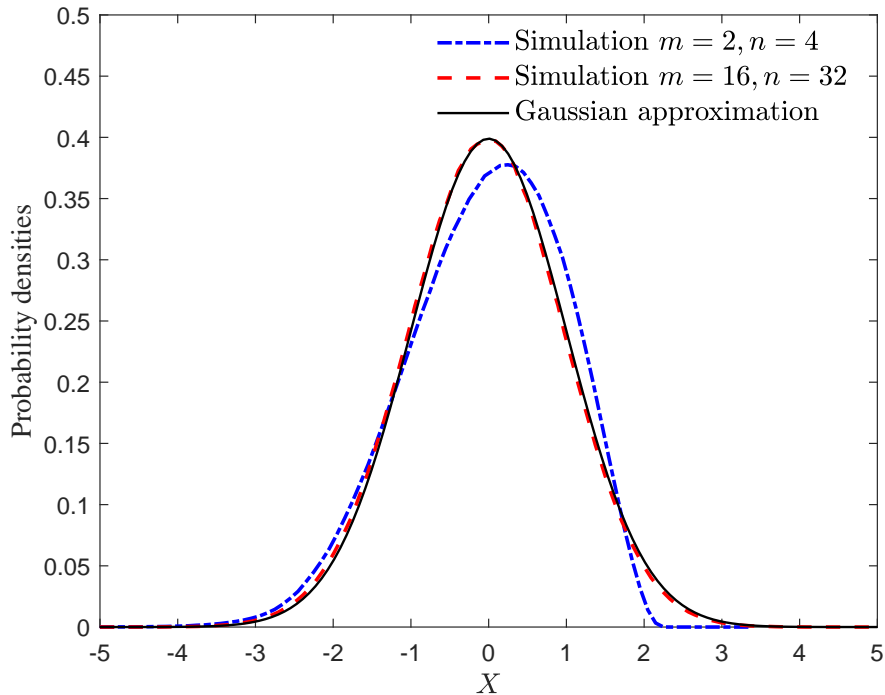}

\caption{Probability densities of standardized von Neumann entropy: a comparison of Gaussian approximation~\eref{eq:iappr} to the simulation results.  The dashed line in blue and the dash-dot line in red refer to the standardized von Neumann entropy~\eref{eq:X} of subsystem dimensions $m=2$, $n=4$, and $m=16$, $n=32$, respectively. The solid black line represents the Gaussian approximation~\eref{eq:iappr}.}
\label{fig:p2}
\end{figure}
In~\fref{fig:p2}, we plot the standardized von Neumann entropy $X$ for the cases of a fixed subsystem dimension ratio $m/n=1/2$ along with the standard Gaussian density
\begin{equation}\label{eq:iappr}
\frac{1}{\sqrt{2\pi}}\e^{-\frac{1}{2}x^{2}}, \qquad x\in\left(-\infty,\infty\right).
\end{equation}
We observe from this plot that the true distribution of the standardized von Neumann entropy $X$ is non-symmetric and appears to be left-skewed when $m=2$, $n=4$. In comparison, when the subsystem dimensions increase to $m=16$, $n=32$ with the same ratio $m/n=1/2$, the distribution of $X$ appears to be closer to the Gaussian distribution. The observed asymptotic Gaussian behavior is typical for a wide class of linear spectral statistics (such as the von Neumann entropy) over different
random matrix ensembles as already been observed in the Hilbert-Schmidt ensemble~\cite{Wei20} and Bures-Hall ensemble~\cite{Wei20BH}. Here, we also conjecture that in the limit
\begin{equation}\label{eq:lim}
m\to\infty,~~~~n\to\infty,~~~~\frac{m}{n}\in(0,1],
\end{equation}
the standardized von Neumann entropy~\eref{eq:X} converges in distribution to a standard Gaussian random variable.
\pagebreak

\begin{figure}[!h]
\centering
\includegraphics[width=0.85\linewidth]{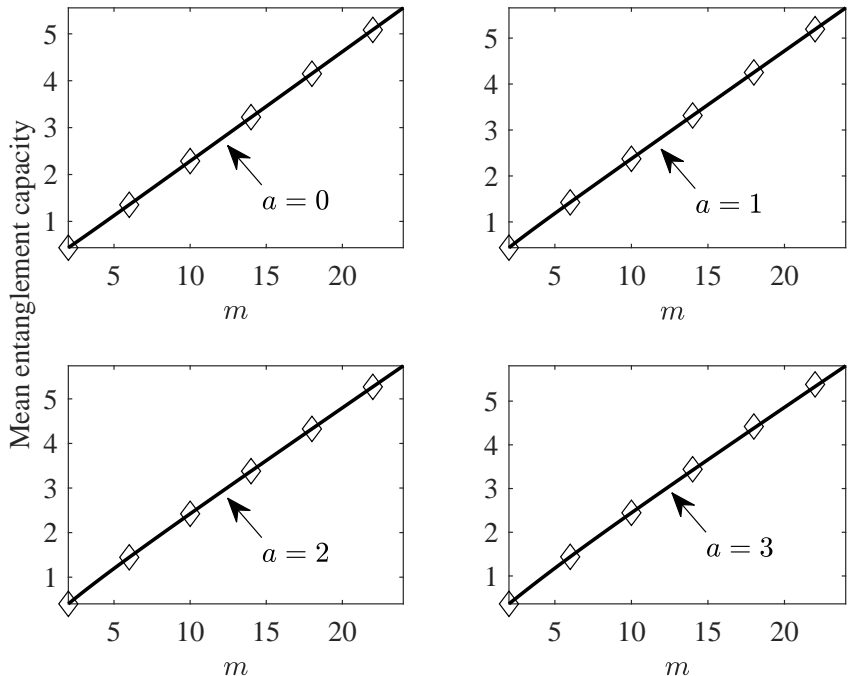}

\caption{Average entanglement capacity: analytical results versus simulations. The solid black lines represent the analytical results~\eref{eq:prop20}--\eref{eq:prop23} and the diamond scatters represent the corresponding simulations. The parameter $a$ in the plots denotes the dimension difference $a=n-m$.}
\label{fig:p3}
\end{figure}

For the entanglement capacity, we numerically compare our mean results~\eref{eq:prop20}--\eref{eq:prop23} with the simulations in~\fref{fig:p3}. As can be seen, the claimed results match well with the simulations. In addition, we notice the linear growth of the mean capacity, which can be captured by the following asymptotic results. For example, when $a=n-m=0$, by taking the appropriate limit of the result~\eref{eq:prop20} and using the asymptotic behavior~\eref{eq:limpl1} of the trigamma function, one obtains the slope of linear behavior
\begin{equation}
\lim_{n \rightarrow \infty}\frac{\mathbb{E}\!\left[C\right]}{n}=\frac{1}{8} \left(\pi ^2-8\right) .
\end{equation}

The rest of the  paper is organized as follows. In~\sref{sec:2}, we perform the derivation of the considered second-order statistics of fermionic Gaussian states. Specifically, in~\sref{sec:2.1} we first provide necessary results of Jacobi polynomials together with two integral identities useful in the subsequent computations. In~\sref{sec:2.2}, we compute the exact variance of von Neumann entropy based on the computation and simplification of the four integrals involved. In~\sref{sec:2.3}, the exact mean capacity is derived by calculating one additional integral along with the results already obtained. The summation forms of the five integrals are listed in appendix A, 
and the relevant finite sum identities utilized in simplifying the summations are listed in appendix B.
\section{Computation of second-order statistics}\label{sec:2}
In this section, we first introduce some useful results of Jacobi polynomials in~\sref{sec:2.1}.  We then present the computation of the variance of von Neumann entropy $\mathbb{V}\!\left[S\right]$ in~\sref{sec:2.2} and that of the mean capacity $\mathbb{E}\!\left[C\right]$ in~\sref{sec:2.3}.
\subsection{Integrals involving Jacobi polynomials}\label{sec:2.1}
We list here the necessary random matrix results and integral identities regarding the Jacobi unitary ensemble. While most of the listed results are known in the literature \cite{Forrester,EMOT-II_1954, Szego}, some of which seem new.

For the considered Jacobi ensemble~(\ref{eq:nJ}), its joint density $g_{l}(x_{1},\dots,x_{l})$ of $l$ (out of $m$) eigenvalues can be written in terms of an $l\times l$ determinant as~\cite{BHK21}
\begin{equation}\label{eq:corrk}
g_{l}(x_{1},\dots,x_{l})=\frac{(m-l)!}{m!}\det\left(K\left(x_{i},x_{j}\right)\right)_{i,j=1}^{l}.
\end{equation}
The corresponding correlation kernel is given by
\begin{equation}\label{eq:kx}
K\left(x,y\right)=\sqrt{(1-x^{2})^{a}(1-y^{2})^{a}}\sum_{k=0}^{m-1}\frac{p_{k}(x)p_{k}(y)}{h_{k}},
\end{equation}
where $a$ has been defined in~(\ref{eq:a}). The polynomial $p_{k}(x)$ of degree $2k$ is related to different representations of the standard Jacobi polynomial~\cite{BHK21,Forrester}
\begin{eqnarray}
\label{eq:J1}\fl J^{(a,b)}_{k}(x)&=&\frac{(-1)^{k}(b+1)_{k}}{k!}\sum_{i=0}^{k}\frac{(-k)_{i}(k+a+b+1)_{i}}{(b+1)_{i}\Gamma(i+1)}\left(\frac{1+x}{2}\right)^{i}\\\label{eq:J2}
\fl &=&\frac{1}{k!}\sum_{i=0}^{k}\frac{(-k)_{i}(k+a+b+1)_{i}(i+a+1)_{k-i}}{\Gamma(i+1)}\left(\frac{1-x}{2}\right)^{i}\\ \label{eq:J3}
\fl &=&\sum _{i=0}^k \frac{(-1)^i \Gamma (a+k+1) (k+b-i+1)_i}{\Gamma (i+1) \Gamma (a+i+1) \Gamma (k-i+1)}\left(\frac{1-x}{2}\right)^i \left(\frac{1+x}{2}\right)^{k-i}
\end{eqnarray}
supported in $x\in[-1,1]$, as
\begin{equation}\label{eq:rela}
p_{k}(x)=J^{(n-m,n-m)}_{2k}(x)
\end{equation}
with
\begin{equation}\label{eq}
(a)_{n}=\frac{\Gamma(a+n)}{\Gamma(a)}
\end{equation}
denoting the Pochhammer's symbol. The orthogonality relation of the standard Jacobi polynomials
\begin{eqnarray}
&&\int_{-1}^{1}(1-x)^{a}(1+x)^{b}J^{(a,b)}_{k}(x)J^{(a,b)}_{l}(x)\dd x\nonumber\\
=&&\frac{2^{a+b+1}\Gamma(k+a+1)\Gamma(k+b+1)}{k!(2k+a+b+1)\Gamma(k+a+b+1)}\delta_{kl}, \quad \Re\{a,b\}>-1,
\end{eqnarray}
leads to the normalization constant $h_{k}$ of the desired polynomials $p_{k}(x)$,
\begin{equation}
\int_{0}^{1}\left(1-x^{2}\right)^{a}p_{k}(x)p_{l}(x)\dd x=h_{k}\delta_{kl}
\end{equation}
as
\begin{equation}\label{eq:h}
h_{k}=\frac{2^{2a}\Gamma^{2}(2k+a+1)}{(4k+2a+1)\Gamma(2k+2a+1)\Gamma(2k+1)}.
\end{equation}
Note that the polynomials~(\ref{eq:rela}) are even
\begin{equation}\label{eq:even}
p_{k}(-x)=p_{k}(x),
\end{equation}
which is a consequence of the parity property of Jacobi polynomials
\begin{equation}\label{eq}
J^{(a,b)}_{k}(-x)=(-1)^{k}J^{(b,a)}_{k}(x).
\end{equation}
We now present two integral identities useful in computing the considered second-order statistics. The first one is
\begin{eqnarray}
&&\int_{-1}^{1}\left(\frac{1-x}{2}\right)^{a}\left(\frac{1+x}{2}\right)^{c}J^{(a,b)}_{k}(x)\dd x \label{eq:Iac}\\
&=&\frac{2\Gamma(c+1)\Gamma(k+a+1)\Gamma(c-b+1)}{k!\Gamma(k+a+c+2)\Gamma(c-k-b+1)}, \quad \Re\{a,b,c\}>-1.\label{eq:SIac}
\end{eqnarray}
This identity is obtained by using the definition~(\ref{eq:J1}) and the identity
\begin{equation}\label{eq1c}
\int_{-1}^{1}(1-x)^{a}(1+x)^{b}\dd x=\frac{2^{a+b+1}\Gamma(a+1)\Gamma(b+1)}{\Gamma(a+b+2)}, \quad \Re\{a,b\}>-1,
\end{equation}
where the resulting summation can be written as a unit argument hypergeometric function of Saalsch\"{u}tzian type that terminates
\begin{equation}
_{3}F_{2}(-n,a,b;d,a+b-n-d+1;1)=\frac{(d-a)_{n}(d-b)_{n}}{(d)_{n}(-a-b+d)_{n}}, \quad n\in\mathbb{Z^{+}}.
\end{equation}
The next identity is
\begin{eqnarray}
&&\int_{-1}^{1}\left(\frac{1-x}{2}\right)^{d}\left(\frac{1+x}{2}\right)^{c}J^{(a,b)}_{k}(x)\dd x  \label{eq:Icd}\\
&=&\frac{2 \Gamma (c-b+1) \Gamma (d-a+1)}{\Gamma (c+d+k+2)}\sum _{i=0}^{k} \frac{(-1)^i \Gamma (c+i+1) \Gamma (d-i+k+1)}{\Gamma (i+1) \Gamma (k-i+1)}\times\nonumber\\
&&\frac{1}{ \Gamma (d-a-i+1) \Gamma (c-b+i-k+1)}, \quad \Re\{a,b,c,d\}>-1,\label{eq:SIcd}
\end{eqnarray}
where the integral involves an additional variable $d$ as compared to the integral~\eref{eq:Iac}.
Proving the above identity requires the Rodrigues' formula\cite{Szego}, i.e., for $\Re\{a,b\}>-1$, one has
\begin{eqnarray}\label{RoF}
\!\!\!\!\!\!\!\!\!\!\!\!\!\!\! (1-x)^{a}(1+x)^{b}J^{(a,b)}_{k}(x)=\frac{(-1)^k}{2^k k!}\frac{\dd^k}{\dd x^k}\left((1-x)^{k+a}(1+x)^{k+b}\right).
\end{eqnarray}
By using the above formula, the integral~\eref{eq:Icd} is evaluated as
\begin{eqnarray}
&& \int_{-1}^{1}\left(\frac{1-x}{2}\right)^{d}\left(\frac{1+x}{2}\right)^{c}J^{(a,b)}_{k}(x)\dd x \nonumber\\
&=&\frac{(-1)^k}{2^{k+c+d} k!}\int_{-1}^{1}(1-x)^{d-a}(1+x)^{c-b}\frac{\dd^k}{\dd x^k}\left((1-x)^{k+a}(1+x)^{k+b}\right)\dd x \label{pintcd2}\\
&=&\frac{1}{2^{k+c+d} k!}\int_{-1}^{1}(1-x)^{k+a}(1+x)^{k+b}\frac{\dd^k}{\dd x^k}\left((1-x)^{d-a}(1+x)^{c-b}\right)\dd x \label{pintcd3}\\
&=&\frac{1}{2^{k+c+d} k!}\sum_{i=0}^{k} {{k}\choose{i}} (-1)^i (d-a-i+1)_i (c-b+i-k+1)_{k-i} \times\nonumber\\
&& \int_{-1}^{1}(1-x)^{d-i+k}(1+x)^{c+i} \dd x \label{pintcd4}\\
&=&\frac{2 \Gamma (c-b+1) \Gamma (d-a+1)}{\Gamma (c+d+k+2)}\sum _{i=0}^{k} \frac{(-1)^i \Gamma (c+i+1) \Gamma (d-i+k+1)}{\Gamma (i+1) \Gamma (k-i+1)}\times\nonumber\\
&&\frac{1}{ \Gamma (d-a-i+1) \Gamma (c-b+i-k+1)}, \quad \Re\{a,b,c,d\}>-1,\label{pintcd5}
\end{eqnarray}
where the equality~\eref{pintcd3} is obtained by applying $k$ times integration by parts and the last equality~\eref{pintcd5} is obtained by the identity~\eref{eq1c}. Finally, we note that besides the results~\eref{eq:SIac} and~\eref{eq:SIcd}, the integrals~\eref{eq:Iac} and~\eref{eq:Icd} also have alternative forms obtained by using the different representations of the Jacobi polynomials such as~\eref{eq:J1}--\eref{eq:J3}. Among the alternative formulas, no other combinations would lead to the desired results in proposition~\ref{prop1} and proposition~\ref{prop2}.

\subsection{Computation of von Neumann entropy variance}\label{sec:2.2}
In this section, we compute the exact variance of von Neumann entropy shown in proposition \ref{prop1}.
The eigenvalue densities required to compute the first two moments of von Neumann entropy can be read off from the $l$-point correlation function~\eref{eq:corrk} as
\begin{eqnarray}
g_{1}(x_{1}) &=& \frac{1}{m}K\left(x_{1},x_{1}\right) \label{eq:1-pt}\\
\!\!\!\!\!\!\!\!g_{2}(x_{1},x_{2}) &=&\frac{1}{m(m-1)}\left(K(x_{1},x_{1})K(x_{2},x_{2})-K^{2}(x_{1},x_{2})\right).
\end{eqnarray}
We now first reproduce the mean formula~\eref{eq:m1} of von Neumann entropy in order to set up our notations while providing a more detailed derivation. Calculating the mean value requires the one-point density~\eref{eq:1-pt} as
\begin{eqnarray}
\mathbb{E}\!\left[S\right]&=&-\int_{0}^{1}v(x)K(x,x)\dd x \label{eq:0m1}.
\end{eqnarray}
 Inserting the definitions~(\ref{eq:vx}) and~(\ref{eq:kx}) into~(\ref{eq:0m1}), one has
\begin{equation}\label{eq:m1s}
\mathbb{E}\!\left[S\right]=-\sum_{k=0}^{m-1}\frac{1}{h_{k}}\int_{-1}^{1}\frac{1+x}{2}\ln\frac{1+x}{2}\left(1-x^{2}\right)^{a}p_{k}^{2}(x)\dd x,
\end{equation}
where $a=n-m$ and it becomes clear that the key is to evaluate the integral
\begin{equation}\label{eq:int0}
\int_{-1}^{1}\left(\frac{1-x}{2}\right)^{a}\left(\frac{1+x}{2}\right)^{c}p_{k}^{2}(x)\dd x.
\end{equation}
By using the identity~\eref{eq:SIac} along with the definitions~(\ref{eq:J1}) and~(\ref{eq:rela}), the above integral is evaluated to
\begin{eqnarray}\label{eq:int1}
&&\int_{-1}^{1}\left(\frac{1-x}{2}\right)^{a}\left(\frac{1+x}{2}\right)^{c}p_{k}^{2}(x)\dd x \nonumber \\
&=&\frac{2\Gamma^{2}(2k+a+1)}{\Gamma(2k+1)\Gamma(2k+2a+1)}\sum_{i=0}^{2k}\frac{(-1)^{i}\Gamma(i+2k+2a+1)}{i!\Gamma(i+a+1)\Gamma(2k-i+1)}\times\nonumber\\
&&\frac{\Gamma(i+c+1)\Gamma(i-a+c+1)}{\Gamma(i-2k-a+c+1)\Gamma(i+2k+a+c+2)}.
\end{eqnarray}
Now taking the derivative of~(\ref{eq:int1}) before setting $c=a+1$, the resulting sum can be evaluated that leads to
\begin{eqnarray}\label{eq:intm}
&&\int_{-1}^{1}\frac{1+x}{2}\ln\frac{1+x}{2}\left(1-x^{2}\right)^{a}p_{k}^{2}(x)\dd x \nonumber\\
&=&h_{k}\big(1+\psi_{0}(2k+a)+\psi_{0}(2k+2a)-2\psi_{0}(4k+2a)\big)+\nonumber\\
&&\frac{h_{k}}{2}\left(\frac{1}{k+a}-\frac{a}{2k+a}-\frac{a}{2k+a+1}-\frac{2}{4k+2a+1}\right),
\end{eqnarray}
where $h_{k}$ is given by~(\ref{eq:h}). Inserting the above result into~(\ref{eq:m1s}), the remaining summation over $k$ is evaluated by repeated use of the identity~\eref{eq:B1} as well as the result~\cite{AS72}
\begin{eqnarray}\label{eq:m_poly0}
\psi_{0}(mk)=\ln m+\frac{1}{m}\sum_{i=0}^{m-1}\psi_{0}\left(k+\frac{i}{m}\right), \qquad m\in\mathbb{Z^{+}}.
\end{eqnarray}
This completes the proof of the mean entropy formula~(\ref{eq:m1}).

We now turn to the variance. By definition, the variance calculation boils down to computing two integrals, cf.~\cite{Wei17,Wei20BH, BHK21},
\begin{eqnarray}\label{def:vs}
\mathbb{V}\!\left[S\right]&=&\mathbb{E}\!\left[S^{2}\right]-\mathbb{E}^{2}\!\left[S\right]=\mathrm{I_{A}}-\mathrm{I_{B}},
\end{eqnarray}
where
\begin{eqnarray}
\mathrm{I_{A}}&=&\int_{0}^{1}v^{2}(x)K(x,x)\dd x \label{eq:IA}\\
\mathrm{I_{B}}&=&\int_{0}^{1}\int_{0}^{1}v(x)v(y)K^{2}\left(x,y\right)\dd x\dd y. \label{eq:IB}
\end{eqnarray}
According to the definitions~\eref{eq:vx} and~\eref{eq:kx}, the $\mathrm{I_{A}}$ integral~(\ref{eq:IA}) boils down computing the two parts
\begin{equation}\label{eq:IAex}
\mathrm{I_{A}}=\mathrm{A_{1}}+\mathrm{A_{2}},
\end{equation}
where
\begin{eqnarray}
\mathrm{A_{1}}&=&\sum_{k=0}^{m-1}\frac{1}{h_{k}}\int_{-1}^{1}\left(\frac{1+x}{2}\right)^{2}\ln^{2}\frac{1+x}{2}\left(1-x^{2}\right)^{a}p_{k}^{2}(x)\dd x\\
\mathrm{A_{2}}&=&\sum_{k=0}^{m-1}\frac{1}{h_{k}}\int_{-1}^{1}\frac{1-x}{2}\frac{1+x}{2}\ln\frac{1-x}{2}\ln\frac{1+x}{2}\left(1-x^{2}\right)^{a}p_{k}^{2}(x)\dd x.
\end{eqnarray}
Similarly, the $\mathrm{I_{B}}$ integral~(\ref{eq:IB}) can be written in terms of the following two integrals
\begin{equation}\label{eq}
\mathrm{I_{B}}=\mathrm{B_{1}}+\mathrm{B_{2}},
\end{equation}
where
\begin{eqnarray}
\!\!\!\!\!\!\!\!\!\mathrm{B_{1}}&=&\sum_{k=0}^{m-1}\frac{1}{h_{k}^{2}}\left(\int_{-1}^{1}\frac{1+x}{2}\ln\frac{1+x}{2}\left(1-x^{2}\right)^{a}p_{k}^{2}(x)\dd x\right)^{2}\\
\!\!\!\!\!\!\!\!\!\mathrm{B_{2}}&=&\sum_{j=1}^{m-1}\sum_{k=0}^{m-j-1}\frac{2}{h_{k+j}h_{k}}\Bigg(\int_{-1}^{1}\frac{1+x}{2}\ln\frac{1+x}{2}\left(1-x^{2}\right)^{a}p_{k+j}(x)p_{k}(x)\dd x\Bigg)^{2}.
\end{eqnarray}

The next step is to compute the integrals in $\mathrm{A_1}$, $\mathrm{A_2}$, $\mathrm{B_1}$, and $\mathrm{B_2}$. These integrals can be calculated into summations by using the corresponding integral identities of Jacobi polynomials introduced in~\sref{sec:2.1}. The strategies in computing the integrals will be discussed in the following. The integral in $\mathrm{A_{1}}$ is calculated by taking twice derivatives of~(\ref{eq:int1}) with respect to $c$ before setting $c=a+2$. The calculation of the integral in $\mathrm{A_{2}}$ requires taking derivatives with respect to both $c$ and $d$ of the identity
\begin{eqnarray}\label{eq:int2}
&&\int_{-1}^{1}\left(\frac{1-x}{2}\right)^{d}\left(\frac{1+x}{2}\right)^{c}p_{k}^{2}(x)\dd x \nonumber \\
&=&\frac{2 \Gamma (2 k+a+1)^2}{\Gamma (4 k+c+d+2)}\sum _{j=0}^{2 k} \sum _{i=0}^{2 k}\frac{(-1)^{i+j} \Gamma (2 k-j+c-a+1)}{\Gamma (i+1) \Gamma (j+1) \Gamma (2 k-j+a+1)}\times\nonumber\\
&&\frac{\Gamma (j-a+d+1) \Gamma (2 k+i-j+c+1) \Gamma (2 k+d-i+j+1)}{\Gamma (j+a+1) \Gamma (2 k-i+1) \Gamma (2 k-j+1) \Gamma (i-j-a+c+1) }\times\nonumber\\
&&\frac{1}{\Gamma (j-i-a+d+1)}
\end{eqnarray}
before setting $c=d=a+1$.
The above identity is established by using the identity~\eref{eq:SIcd} along with the definitions~(\ref{eq:J3}) and~(\ref{eq:rela}).
The integral in~$\mathrm{B_{1}}$ is given by~(\ref{eq:intm}), which has been obtained in the mean calculation. The integral in~$\mathrm{B_{2}}$ is computed by taking the derivative of the identity
\begin{eqnarray}\label{eq:int3}
&&\int_{-1}^{1}\left(\frac{1-x}{2}\right)^{a}\left(\frac{1+x}{2}\right)^{c}p_{k+j}(x)p_{k}(x)\dd x \nonumber \\
&=&\frac{2\Gamma(2k+a+1)\Gamma(2k+2j+a+1)}{\Gamma(2k+2j+1)\Gamma(2k+2a+1)}\sum_{i=0}^{2k}\frac{(-1)^{i}\Gamma(i+2k+2a+1)}{\Gamma(i+a+1)\Gamma(2k-i+1)}\times\nonumber\\
&&\frac{\Gamma(i+c+1)\Gamma(i-a+c+1)}{\Gamma(i+1)\Gamma(i-2k-2j-a+c+1)\Gamma(i+2k+2j+a+c+2)}
\end{eqnarray}
with respect to $c$ and setting $c=a+1$. To derive the above identity, we first apply the definitions~\eref{eq:J1} and~\eref{eq:J2} on $p_{k+j}(x)$ and $p_{k}(x)$, respectively. Keeping in mind the relation~\eref{eq:rela}, we have
\begin{eqnarray}\label{eq:pint3}
&&\int_{-1}^{1}\left(\frac{1-x}{2}\right)^{a}\left(\frac{1+x}{2}\right)^{c}p_{k+j}(x)p_{k}(x)\dd x \nonumber \\
&=&\frac{\Gamma (a+2 k+1) \Gamma (a+2 j+2 k+1)}{\Gamma (2 a+2 k+1) \Gamma (2 a+2 j+2 k+1)}\sum _{s=0}^{2 j+2 k} \frac{(-1)^s }{\Gamma (s+1) \Gamma (a+s+1) }\times\nonumber\\
&&\frac{\Gamma (2 a+2 j+2 k+s+1)}{\Gamma (2 j+2 k-s+1)}\sum _{i=0}^{2 k} \frac{(-1)^i \Gamma (2 a+i+2 k+1)}{\Gamma(i+1) \Gamma (a+i+1) \Gamma (2k-i+1)}\times\nonumber\\
&&\int_{-1}^1 \left(\frac{1-x}{2}\right)^{a+s} \left(\frac{1+x}{2}\right)^{c+i} \dd x. \label{eq:pint3_1}
\end{eqnarray}
The integral in~\eref{eq:pint3_1} is further evaluated by the identity~\eref{eq1c}, where one of the resulting summations can be written as a unit argument $_2F_1$ hypergeometric function
\begin{eqnarray}
 && _2F_1(-2 j-2 k,2 a+2 j+2 k+1;a+c+i+2;1)\nonumber \\
&=&\frac{\Gamma (-a+c+i+1) \Gamma (a+c+i+2)}{\Gamma (-a+c+i-2 j-2 k+1) \Gamma (a+c+i+2 j+2 k+2)}
\end{eqnarray}
leading to the claimed identity~\eref{eq:int3}.
In writing down the summation forms of $\mathrm{A_1}$, $\mathrm{A_2}$, $\mathrm{B_1}$, and $\mathrm{B_2}$, one will have to solve the indeterminacy by using the following formulas for $\epsilon \rightarrow 0$,
\numparts
\label{eq:pgna}\begin{eqnarray}
\Gamma(-l+\epsilon)&\!=&\frac{(-1)^{l}}{l!\epsilon}\left(1+\psi_{0}(l+1)\epsilon+o\left(\epsilon^2\right)\right)\label{eq:pgna1}\\
\psi_{0}(-l+\epsilon)&=&-\frac{1}{\epsilon}+\psi_{0}(l+1)+\left(2\psi_{1}(1)-\psi_{1}(l+1)\right)\epsilon+o\left(\epsilon^2\right)\label{eq:pgna2}\\
\psi_{1}(-l+\epsilon)&=&\frac{1}{\epsilon^{2}}-\psi_{1}(l+1)+\psi_{1}(1)+\zeta(2)+o\left(\epsilon\right)\label{eq:pgna3}
\end{eqnarray}
\endnumparts
with $l$ being a non-positive integer. The resulting summations forms of $\mathrm{A_1}$, $\mathrm{A_2}$, $\mathrm{B_1}$, and $\mathrm{B_2}$ are summarized in appendix A. The rest of the task to show proposition~\ref{prop1} is to simplify these summations.

The formulas in appendix A involve summations over gamma, digamma, and trigamma functions. The identities utilized in evaluating these sums are listed in appendix B, where the closed-form identities are~\eref{eq:B1}--\eref{eq:B5} and the semi closed-form ones are~\eref{eq:B6}--\eref{eq:B8}. These semi closed-form identities represent the relation between two single sums consisting of rational functions and polygamma functions. By using these semi closed-form identities, one is able to convert different single sums into the ones that we refer to as unsimplifiable basis such as the sums~\eref{eq:bs1}--\eref{eq:bs3}.  Note that most of the summations in appendix A do not admit a closed-form evaluation individually. To facilitate the cancellations among the summations, one has to potentially first simplify these sums into semi closed-form ones via the unsimplifiable basis. The closed-form results in this work become available by observing the cancellations among the semi closed-form terms.

The simplification process of the summation forms~\eref{eq:A1S}--\eref{eq:B2S} in appendix A is outlined below.
For an arbitrary $a$, the summation forms~\eref{eq:A1S} and~\eref{eq:B1S} only involve rational functions. Thus, these sums admit semi closed-form expressions by using the identities listed in appendix B along with the results~\eref{eq:m_poly0} and
\begin{eqnarray}\label{eq:m_poly1}
\psi_{1}(mk)=\frac{1}{m^2}\sum _{i=0}^{m-1} \psi _1\left(\frac{i}{m}+k\right), \qquad m\in\mathbb{Z^{+}}.
\end{eqnarray}
The summations~\eref{eq:A2S} and~\eref{eq:B2S} may not be represented via the semi closed-form terms. This fact naturally guides us to first consider the special case of equal subsystem dimensions $a=n-m=0$, where the summations can be presented in terms of semi-closed expressions. Specifically, the summation representation~\eref{eq:A2S} is simplified by first using the identity
\begin{eqnarray}\label{eq:id1}
\!\!\!&&\sum _{k=0}^m \frac{1}{\Gamma (k+1) \Gamma (c+k) \Gamma (m+1-k) \Gamma (m+b+1-k)}\\
\!\!\!&=&\frac{\, _2F_1(-b-m,-m;c;1)}{\Gamma (c) \Gamma (m+1) \Gamma (b+m+1)}\nonumber\\
\!\!\!&=&\frac{\Gamma (b+c+2 m)}{\Gamma (m+1) \Gamma (b+m+1) \Gamma (c+m) \Gamma (b+c+m)}, \quad \Re(2m+c+b)>0
\end{eqnarray}
and identities obtained by up to its second derivatives of $c$ and $b$,
\begin{eqnarray}\label{eq:id1_1}
\fl  &&\sum _{k=0}^m \frac{\psi _{0}(m+b+1-k)}{\Gamma (k+1) \Gamma (c+k) \Gamma (m+1-k) \Gamma (m+b+1-k)}\nonumber \\
\fl &=&\frac{\Gamma (b+c+2 m) (\psi_{0}(b+c+m)-\psi_{0}(b+c+2 m)+\psi_{0}(b+m+1))}{\Gamma (m+1) \Gamma (b+m+1) \Gamma (c+m) \Gamma (b+c+m)}
\end{eqnarray}
\begin{eqnarray}\label{eq:id1_2}
 \fl &&\sum _{k=0}^m \frac{\psi _{0}(c+k)}{\Gamma (k+1) \Gamma (c+k) \Gamma (m+1-k) \Gamma (m+b+1-k)}\nonumber \\
\fl &=&\frac{\Gamma (b+c+2 m) (\psi_{0}(b+c+m)-\psi_{0}(b+c+2 m)+\psi_{0}(c+m))}{\Gamma (m+1) \Gamma (b+m+1) \Gamma (c+m) \Gamma (b+c+m)}
\end{eqnarray}
\begin{eqnarray}\label{eq:id1_3}
\fl  &&\sum _{k=0}^m \frac{\psi _{0}(c+k)\psi _{0}(m+b+1-k)}{\Gamma (k+1) \Gamma (c+k) \Gamma (m+1-k) \Gamma (m+b+1-k)}\nonumber \\
\fl &=&\frac{\Gamma (b+c+2 m) }{\Gamma (m+1) \Gamma (b+m+1) \Gamma (c+m) \Gamma (b+c+m)}((\psi_{0}(b+c+m)-\psi_{0}(b+c+2 m)+\nonumber\\
\fl &&\psi_{0}(b+m+1)) (\psi_{0}(b+c+m)-\psi_{0}(b+c+2 m)+\psi_{0}(c+m))-\psi_{1}(b+c+m)+\nonumber\\
\fl &&\psi_{1}(b+c+2 m)).
\end{eqnarray}
All the gamma functions cancel in pairs in the resulting summations, where the remaining terms can be evaluated by the identities in appendix B that leads to a semi closed-form representation. For the summation~\eref{eq:B2S}, it directly reduces to a sum of rational functions when $a=0$. This also leads to a semi closed-form representation of~\eref{eq:B2S}. With the above results, the integrals $\mathrm{I_A}$ and $\mathrm{I_B}$ can now be obtained. For $\mathrm{I_A}$, one has
\begin{eqnarray}\label{eq:IAa0}
\fl \mathrm{I_A}=&\sum _{k=1}^n \left(\frac{\psi _0(2 k)}{k}-\frac{\psi _0(4 k)}{2 k}+\frac{\psi _0(4 k)}{2 k+1}\right)-\frac{24 n^2-12 n+1}{4 (4 n-1)}\psi _1(2 n)-\frac{1}{4}\psi _1(n)+\nonumber\\
\fl &(4 n-1) \psi _0^2(4 n)+2 (1-4 n) \psi _0(2 n) \psi _0(4 n)+\frac{1}{2} (8 n-3)\psi _0^2(2 n)+\psi _0(n) \psi _0(2 n)-\nonumber\\
\fl &\frac{1}{2} \psi _0^2(n)-\frac{16 n^3+8 n^2-1}{2 n (2 n+1)}\psi _0(4 n)+\frac{1}{2} (8 n+1)\psi _0(2 n)-\left(\frac{1}{2 n}+\ln 2\right)\psi _0(n)-\nonumber\\
\fl &\frac{1}{2} \psi _0(n) \psi _0\!\left(n+\frac{1}{2}\right)-\frac{2 n+1}{2 n}\psi _0\!\left(n+\frac{1}{2}\right)+\frac{1}{2} \psi _0\!\left(n+\frac{1}{4}\right)+\frac{n (5 n-2) }{4 n-1}\psi _1(1)+\nonumber\\
\fl &\!\left(\frac{1}{2}+\ln 2\right)\psi _0(1)+\frac{1}{2} \psi _0\!\left(\frac{1}{2}\right) \psi _0(1)+\psi _0\!\left(\frac{1}{2}\right)-\frac{1}{2} \psi _0\!\left(\frac{1}{4}\right)-\frac{1}{n}\ln 2-n+2.
\end{eqnarray}
Similarly, for $\mathrm{I_B}$, one arrives at
\begin{eqnarray}\label{eq:IBa0}
\fl \mathrm{I_B}=&\sum _{k=1}^n \left(\frac{\psi _0(2 k)}{k}-\frac{\psi _0(4 k)}{2 k}+\frac{\psi _0(4 k)}{2 k+1}\right)+\frac{4 n-1}{2}  \psi _1(4 n)-\frac{104 n^2-60 n+7}{8 (4 n-1)}\psi _1(2 n)-\nonumber\\
\fl&\frac{3}{8} \psi _1(n)+(4 n-1) \psi _0^2(4 n)-2 (4n-1) \psi _0(2 n) \psi _0(4 n)+\frac{1}{2} (8 n-3) \psi _0^2(2 n)+\psi _0(n)\times\nonumber\\
\fl & \psi _0(2 n)-\frac{1}{2} \psi _0^2(n)+\frac{-16 n^3-6 n^2+n+1}{2 n (2 n+1)}\psi _0(4 n)+4 n\psi _0(2 n)-\left(\frac{1}{2 n}+\ln 2\right)\times\nonumber\\
\fl &\psi _0(n) - \frac{1}{2} \psi _0(n) \psi _0\!\left(n+\frac{1}{2}\right)-\frac{(2 n+1) }{2 n}\psi _0\!\left(n+\frac{1}{2}\right)+\frac{1}{2} \psi _0\!\left(n+\frac{1}{4}\right)+\frac{ (5 n-2) }{4 n-1}\times\nonumber\\
\fl &n \psi _1(1)+ \left(\frac{1}{2}+\ln 2\right)\psi _0(1)+\frac{1}{2} \psi _0\!\left(\frac{1}{2}\right) \psi _0(1)-\frac{1}{2} \psi _0\!\left(\frac{1}{4}\right)+\psi _0\!\left(\frac{1}{2}\right)-\nonumber\\
\fl &\frac{1}{n}\ln 2-n+2.
\end{eqnarray}
Now inserting the $\mathrm{I_A}$ expression~\eref{eq:IAa0} and the above $\mathrm{I_B}$ expression~\eref{eq:IBa0} into~\eref{def:vs}, we observe substantial cancellations among the terms of $\mathrm{I_A}-\mathrm{I_B}$. In particular, the terms involving the three types of sums
\numparts
\begin{eqnarray}
\sum _{k=1}^n \frac{\psi _0(2 k)}{k}\label{eq:bs1}\\
\sum _{k=1}^n \frac{\psi _0(4 k)}{k}\label{eq:bs2}\\
\sum _{k=1}^n \frac{\psi _0(4 k)}{2 k+1}\label{eq:bs3}
\end{eqnarray}
\endnumparts
cancel completely. The surviving terms give us the desired identity~\eref{eq:prop1}. This completes the proof of proposition \ref{prop1}. By using the same approach as for the equal subsystem dimension case $a=0$, we also compute the variance for cases being $a=1,2,3$. Note that cancellation phenomenon also appears in other statistic of entanglement measures, see for example~\cite{Wei17,Wei20BH,Wei20,HWC21}.
\subsection{Computation of mean capacity}\label{sec:2.3}
In this section, we compute the average capacity of entanglement in fermionic Gaussian states. By definitions~\eref{eq:capacity} and~\eref{eq:corrk}, computing the average capacity requires only the one-point density~\eref{eq:1-pt} as
\begin{eqnarray}\label{eq:Ca0}
\mathbb{E}\!\left[C\right]=\mathrm{I_C}-\mathrm{I_A},
\end{eqnarray}
where
\begin{eqnarray}
\mathrm{I_C}&=&\int_{0}^{1}\left(\frac{1+x}{2}\ln^{2}\left(\frac{1+x}{2}\right)+\frac{1-x}{2}\ln^{2}\left(\frac{1-x}{2}\right)\right)K(x,x)\dd x,
\end{eqnarray}
and $\mathrm{I_A}$ has been defined in~\eref{eq:IAex}.
Using the correlation kernel~\eref{eq:kx} of one-point density, the integral $\mathrm{I_C}$ can be written as
\begin{eqnarray}
\mathrm{I_C}=\sum_{k=0}^{m-1}\frac{1}{h_{k}}\int_{-1}^{1}\frac{1+x}{2}\ln^{2}\left(\frac{1+x}{2}\right)\left(1-x^{2}\right)^{a}p_{k}^{2}(x)\dd x.
\end{eqnarray}
The above integral can be evaluated by taking twice derivatives of the identity~\eref{eq:int1} with respect to $c$ before setting $c=a+1$.  The resulting summation of $\mathrm{I_C}$ is given by~\eref{eq:ICS} in appendix A. Similar to $\mathrm{A_1}$, $\mathrm{I_C}$ also admits a semi-closed expression for any $a$. To proceed further, we have to consider the case when $a=0$ due to the intractability of $\mathrm{I_A}$ as discussed previously. Indeed, when $a=0$, we immediately have
\begin{eqnarray}\label{eq:ICa0}
\fl \mathrm{I_C}=&\sum _{k=1}^n \left(\frac{\psi _0(2 k)}{k}-\frac{\psi _0(4 k)}{2 k}+\frac{\psi _0(4 k)}{2 k+1}\right)-\frac{1}{4} (8 n-3) \psi _1(2 n)-\frac{3}{8} \psi _1(n)+(4 n-1) \times\nonumber\\
\fl &\psi _0^2(4 n)-2 (4 n-1) \psi _0(2 n) \psi _0(4 n)+\frac{1}{2} (8 n-3) \psi _0^2(2 n)-\frac{16 n^3+8 n^2-1}{2 n (2 n+1)}\psi _0(4 n)+\nonumber\\
\fl &\frac{\left(8 n^2-3 n-2\right) }{2 n}\psi _0(2 n)+\psi _0(n)+\frac{1}{2} \psi _0\!\left(n+\frac{1}{4}\right)+\frac{1}{8} (16 n-3) \psi _1(1)+\frac{1}{2} \psi _0^2(1)+\nonumber\\
\fl &\frac{3}{2} \psi _0(1)-\frac{1}{2} \psi _0\!\left(\frac{1}{4}\right)-2 n+\frac{5}{2}.
\end{eqnarray}
By inserting the $\mathrm{I_A}$ expression~\eref{eq:IAa0} and the above $\mathrm{I_C}$ expression~\eref{eq:ICa0} into~\eref{eq:Ca0}, we arrive at the claimed result~\eref{eq:prop20} of the case $a=0$ after observing the cancellations. In the same manner, the capacity formulas~\eref{eq:prop21}--\eref{eq:prop23} that correspond to the cases $a=1,2,3$ are obtained.

\section{Conclusion}
In this work, we compute the two major second-order statistics of fermionic Gaussian states -- the variance of von Neumann entanglement entropy and the average entanglement capacity. A key ingredient in obtaining these results is based on representing the underlying integrals into appropriate summations as well as the use of semi closed-form basis leading to a complete cancellation. We also conjectured an explicit expression for the variance of von Neumann entropy valid for any subsystem dimensions.

\section*{Acknowledgments}
The authors wish to thank Lucas Hackl, Santosh Kumar, and Nicholas Witte for correspondence. The work of Lu Wei is supported in part by the U.S. National Science Foundation ($\#$2150486 and $\#$2006612).

\appendix\

\section{Summation representations of integrals in $\mathrm{I_A}$, $\mathrm{I_B}$, and $\mathrm{I_C}$}\label{App_SE}
In this appendix, we list the summation representations of the integrals in $\mathrm{I_A}$, $\mathrm{I_B}$, and $\mathrm{I_C}$.
\begin{eqnarray}\label{eq:A1S}
\fl \mathrm{A_1}=&\sum _{k=0}^{m-1}2(2 a+4 k+1)\left(\rule{0cm}{0.75cm}\sum _{j=2 k-2}^{2 k}\frac{(-1)^j (j+1)_2 (a+j+1)_2}{\Gamma (2k-j+1) \Gamma (j-2 k+3) (2 a+j+2 k+1)_3}\right.\times\nonumber\\
\fl &\bigg((\psi_0 (a+j+3)-\psi_0 (2 a+j+2 k+4)-\psi_0 (j-2 k+3)+\psi_0 (j+3))^2-\nonumber\\
\fl &\psi _1(2 a+j+2 k+4)+\psi _1(a+j+3)-\psi _1(j-2 k+3)+\psi _1(j+3)\bigg)+\nonumber\\
\fl & \sum _{j=0}^{2 k-3} \frac{2 (j+1)_2 (a+j+1)_2}{(2k-j-2)_3 (2 a+j+2 k+1)_3}(\psi_0 (2 a+j+2 k+4)-\psi_0 (a+j+3)+\nonumber\\
\fl &\!\left.\psi_0 (2k-j-2)-\psi_0 (j+3))\rule{0cm}{0.75cm}\right)
\end{eqnarray}

\begin{eqnarray}\label{eq:A2S}
\fl \mathrm{A_2}=&\sum _{k=0}^{m-1} \frac{(2 a+4 k+1) \Gamma (2 k+1) \Gamma (2 a+2 k+1)}{\Gamma (2 a+4 k+4)}\left(\rule{0cm}{0.75cm}\sum _{i=0}^{2 k}\frac{2 (i+1) (2k-i+1) }{\Gamma (i+1)\Gamma (a+i+1)}\times\right.\nonumber\\
\fl &\frac{\Gamma (a+2 k+2)^2}{  \Gamma (2k-i+1) \Gamma (a+2k-i+1)}((\psi_0 (a+2 k+2)-\psi_0 (2 a+4 k+4)-\psi_0 (2)+\nonumber\\
\fl &\psi_0 (2k-i+2)) (\psi_0 (a+2 k+2)-\psi_0 (2 a+4 k+4)+\psi_0 (i+2)-\psi_0 (2))-\nonumber\\
\fl &\psi _1(2 a+4 k+4))-\sum _{j=0}^{2 k} \frac{(j+1) \Gamma (a+2 k+1) \Gamma (a+2 k+3)}{\Gamma (j) \Gamma (a+j+1) \Gamma (2k-j+1)\Gamma (a-j+2 k+1)}\times\nonumber\\
\fl & ((\psi_0 (a+2 k+1)-\psi_0 (2 a+4 k+4)+\psi_0 (2k-j+2)-\psi_0 (1)) (\psi_0 (a+2 k+3)-\nonumber\\
\fl &\psi_0 (2 a+4 k+4)+\psi_0 (j+2)-\psi_0 (3))-\psi _1(2 a+4 k+4))-\sum _{j=0}^{2 k} \frac{(2k-j+1)}{\Gamma (j+1)}\times\nonumber\\
\fl &\frac{ \Gamma (a+2 k+1) \Gamma (a+2 k+3)}{ \Gamma (a+j+1) \Gamma (2 k-j) \Gamma (2 k-j+a+1)}((\psi_0 (a+2 k+3)-\psi_0 (2 a+4 k+4)\nonumber\\
\fl & +\psi_0 (2k-j+2)-\psi_0 (3)) (\psi_0 (a+2 k+1)-\psi_0 (2 a+4 k+4)+\psi_0 (j+2)-\nonumber\\
\fl &\left.\psi_0 (1))-\psi _1(2 a+4 k+4))\rule{0cm}{0.75cm}\right)+\sum _{k=0}^{m-1} \frac{4 (2 a+4 k+1)  \Gamma (2 k+1)\Gamma (2 a+2 k+1)}{\Gamma (2 a+4 k+4)}\times\nonumber\\
\fl & \sum _{j=0}^{2 k} \sum _{i=0}^{2k-j-2}\frac{ (2 k-i-j-1)(i+j+3) \Gamma (a-j+2 k) \Gamma (a+j+2 k+4)}{\Gamma (i+1)  \Gamma (2k-i+1) \Gamma (a+i+j+3) \Gamma (a-i-j+2 k-1)}\times\nonumber\\
\fl &\frac{1}{(j+1)_3}(\psi_0 (a+j+2 k+4)-\psi_0 (2 a+4 k+4)+\psi_0 (i+j+4)-\psi_0 (j+4))
\end{eqnarray}

\begin{eqnarray}\label{eq:B1S}
\fl \mathrm{B_1}=&\sum _{k=0}^{m-1} \left(\psi_0 (a+2 k)+\psi_0 (2 a+2 k)-2 \psi_0 (2 a+4 k)-\frac{1}{2} \left(\frac{a}{a+2 k+1}+\frac{a}{a+2 k}+\right.\right.\nonumber\\
\fl &\!\left.\left.\frac{2}{2 a+4 k+1}\right)+1\right)^2
\end{eqnarray}

\begin{eqnarray}\label{eq:B2S}
\fl \mathrm{B_2}=&\sum _{k=0}^{m-1} \sum _{j=1}^{m-k-1}\frac{\Gamma (2 a+2 k+1) \Gamma (2 j+2 k+1)(2 a+4 k+1) (2 a+4 j+4 k+1)}{2\Gamma (2 k+1) \Gamma (2 a+2 j+2 k+1) j^2 (2 j-1)^2 (2 j+1)^2}\times\nonumber\\
\fl & \frac{ \left(2 a^2 j+a^2+2 a j^2+4 a j k+3 a j+4 a k+a+2 j^2+4 j k+j+4 k^2+2 k\right)^2}{ (a+j+2 k)^2 (a+j+2 k+1)^2 (2 a+2 j+4 k+1)^2}
\end{eqnarray}
\begin{eqnarray}\label{eq:ICS}
\fl \mathrm{I_C}=&\!\left(\psi _0(a+2)-\psi _0(2 a+3)\right)^2+\psi _1(a+2)-\psi _1(2 a+3)+\sum _{k=1}^{m-1} 2 (2 a+4 k+1)\times\nonumber\\
\fl &\left(\rule{0cm}{0.75cm}\sum _{j=2 k-1}^{2 k} \frac{(-1)^j (j+1) (a+j+1)}{(2 a+j+2 k+1)_2}\bigg((\psi _0(j+2)-\psi _0(2 a+j+2 k+3)+\right.\nonumber\\
\fl &\psi _0(a+j+2)-\psi _0(j-2 k+2))^2+\psi _1(a+j+2)-\psi _1(2 a+j+2 k+3)+\nonumber\\
\fl &\psi _1(j+2)-\psi _1(j-2 k+2)\bigg)+\sum _{j=0}^{2 k-2} \frac{2 (j+1) (a+j+1)}{(2k-j-1)_2 (2 a+j+2 k+1)_2} \times\nonumber\\
\fl &\!\left.\left(\psi _0(a+j+2)-\psi _0(2 a+j+2 k+3)-\psi _0(2k-j-1)+\psi _0(j+2)\right)\rule{0cm}{0.75cm}\right)
\end{eqnarray}
\section{List of summation identities}\label{App_SI}
In this appendix, we list the finite sum identities useful in simplifying the summations in appendix A. Except for the identity~\eref{eq:B5} that seems new, the others can be found in~\cite{Wei17,Wei20,Wei20BH,HWC21,Milgram}.
 Here, it is sufficient to set $a,b\ge0, a\neq b$ in identities~\eref{eq:B1}--\eref{eq:B3},~\eref{eq:B6}--\eref{eq:B7}, and $a>m$ in~\eref{eq:B8}. The derivation of the identity~\eref{eq:B5} is provided in the end of this appendix.
\begin{eqnarray}\label{eq:B1}
\fl\sum_{k=1}^{m}\psi_{0}(k+a)=(m+a)\psi_{0}(m+a+1)-a\psi_{0}(a+1)-m
\end{eqnarray}
\begin{equation}\label{eq:B2}
\fl\sum_{k=1}^{m}\psi_{1}(k+a)=(m+a)\psi_{1}(m+a+1)-a\psi_{1}(a+1)+\psi_{0}(m+a+1)-\psi_{0}(a+1)
\end{equation}
\begin{eqnarray}\label{eq:B3}
\fl\sum_{k=1}^{m}\frac{\psi_{0}(k+a)}{k+a}=\frac{1}{2}\left(\psi_{1}(m+a+1)-\psi_{1}(a+1)+\psi_{0}^{2}(m+a+1)-\psi_{0}^{2}(a+1)\right)
\end{eqnarray}
\begin{eqnarray}\label{eq:B4}
\fl\sum_{k=1}^{m}\frac{\psi_{0}(m+1-k)}{k}=\psi_{0}^{2}(m+1)-\psi_{0}(1)\psi_{0}(m+1)+\psi_{1}(m+1) -\psi_{1}(1)
\end{eqnarray}
\begin{eqnarray}\label{eq:B5}
\fl\sum _{k=1}^m \frac{\psi_0 (m+1+k)}{k}=\psi _0^2(m+1)-\psi _0(1) \psi _0(m+1)-\frac{1}{2} \psi _1(m+1)+\frac{\psi _1(1)}{2}
\end{eqnarray}
\begin{eqnarray}\label{eq:B6}
\fl\sum_{k=1}^{m}\psi_{0}(k+a)\psi_{0}(k+b)&=&(b-a) \sum _{k=1}^{m-1} \frac{\psi_0(a+k)}{b+k}+(m+a) \psi_0(m+a) \psi_0(m+b)-a\times\nonumber\\
&& \psi_0(a+1) \psi_0(b+1)-(m+a-1) \psi_0(m+a)+a \psi_0(a+1)-\nonumber\\
&&(m+b) \psi_0(m+b)+(b+1) \psi_0(b+1)+2 m-2
\end{eqnarray}
\begin{eqnarray}\label{eq:B7}
\fl \sum_{k=1}^{m}\frac{\psi_{0}(k+b)}{k+a}=&-\sum_{k=1}^{m}\frac{\psi_{0}(k+a)}{k+b}+\psi_{0}(m+a+1)\psi_{0}(m+b+1)-\psi_{0}(a+1)\times\nonumber\\
&\psi_{0}(b+1)+\frac{1}{a-b}(\psi_{0}(m+a+1)-\psi_{0}(m+b+1)-\psi_{0}(a+1)+\nonumber\\
&\psi_{0}(b+1))
\end{eqnarray}
\begin{eqnarray}\label{eq:B8}
\fl \sum_{k=1}^{m}\frac{\psi_{0}(a+1-k)}{k}=&-\sum_{k=1}^{m}\frac{\psi_{0}(k+a-m)}{k}+(\psi_{0}(a-m)+\psi_0(a+1))(\psi_{0}(m+1)-\nonumber\\
&\psi_{0}(1))+\frac{1}{2}\left((\psi_{0}(a-m)-\psi_{0}(a+1))^2+\psi_{1}(a+1)-\psi_{1}(a-m)\right) \nonumber\\
\end{eqnarray}
The detailed steps in deriving the identity~\eref{eq:B5} are as follows.
\begin{eqnarray}
&&\sum _{k=1}^m \frac{\psi_0 (m+1+k)}{k}\label{eq:B7s0}\\
&=& \sum _{k=1}^m\sum _{l=1}^m \frac{1}{k (k+l)}+\sum _{k=1}^m \frac{\psi_0 (k+1)}{k}\label{eq:B7s1}\\
&=&2 \sum _{k=1}^m \frac{\psi_0(k+1)}{k}+ \sum _{l=1}^m \frac{\psi_0(m+1)-\psi_0(1)}{l}-\sum _{l=1}^m \frac{\psi_0(m+1+l)}{l}\label{eq:B7s2}.
\end{eqnarray}
The equality~\eref{eq:B7s1} is obtained by representing $\psi_0 (m+1+k)$ in terms of $\psi_0 (k+1)$ and $m$ rational terms by using~\eref{eq:pl0}. The equality~\eref{eq:B7s2} is obtained by changing the summation order to first sum over $l$, where one obtains a same summation as in~\eref{eq:B7s0} except for a negative sign. The remaining sums can be simplified that leads to the identity~\eref{eq:B5}.

\end{document}